\documentclass{article}
\usepackage{spconf,amsmath,graphicx,hyperref}
\usepackage{cite}
\usepackage{booktabs}
\usepackage{multirow}
\usepackage{xcolor}
\usepackage{hyperref}
\usepackage{amssymb}

\title{Bridge, Distill, Refine: High-Fidelity One-Step Generation for Multi-Bandwidth Speech Extension}
\name{
    Zhiqi Ai\textsuperscript{1,$\dagger$},\thanks{\textsuperscript{$\dagger$}Work done during Zhiqi’s internship at OPPO}
    Xu Shen\textsuperscript{4},
    Mou Wang\textsuperscript{2,*},
    Yongjin Zhou\textsuperscript{1,*},
    Shugong Xu\textsuperscript{3,*}\thanks{\textsuperscript{*}Corresponding authors}
}
\address{
\textsuperscript{1}Shanghai University, Shanghai, China \\
\textsuperscript{2}OPPO, Beijing \\
\textsuperscript{3}Xi'an Jiaotong Liverpool University, Suzhou, China \\
\textsuperscript{4}Waseda University, Japan \\
}

\begin{document}
\ninept
\maketitle
%
% \begin{abstract}
% Speech bandwidth extension (BWE) aims to recover missing spectral content from band-limited speech. However, most generative BWE methods are designed for fixed input bandwidths and require iterative sampling, limiting their flexibility and efficiency. We propose \textbf{BDR-BWE}, a three-stage generative framework for multiple input bandwidths. First, a cutoff-aware conditional flow bridge is learned in the compressed complex STFT domain. Weak perturbations are applied to observed frequencies and stronger ones to missing frequencies, preserving available information while focusing generation on spectral completion. Second, progressive consistency distillation compresses the multi-step transport into a one-step generator. Third, adversarial refinement aligns generated spectra with real full-band speech, improving high-frequency detail and harmonic coherence. BDR-BWE achieves state-of-the-art log-spectral distance (LSD) on VCTK across all evaluated bandwidths, while maintaining competitive perceptual quality and generalizing well to EARS. Inference requires only a single network evaluation.
% \end{abstract}
\begin{abstract}
Speech bandwidth extension (BWE) aims to recover missing spectral content from band-limited speech. However, most generative BWE methods are designed for fixed input bandwidths and rely on iterative sampling, limiting their flexibility and efficiency. We propose \textbf{BDR-BWE}, a three-stage generative framework for multi-bandwidth speech extension. First, to enable conditional generation from the observed band-limited spectrum, we design a cutoff-aware conditional flow bridge in the compressed complex STFT domain, focusing generation on the missing high-frequency band while preserving observed spectral information. Second, progressive consistency distillation compresses the multi-step flow transport into a one-step generator. Third, adversarial refinement further improves high-frequency detail and harmonic coherence. BDR-BWE achieves state-of-the-art log-spectral distance (LSD) on VCTK across all evaluated bandwidths with one-step inference, while maintaining competitive perceptual quality and generalizing well to EARS. Audio samples are available at \url{https://aizhiqi-work.github.io/bdr-bwe-demo/}.
\end{abstract}

\section{Introduction}
\label{sec:intro}

Speech bandwidth extension (BWE), also known as speech super-resolution,
reconstructs missing frequency components from band-limited speech. It is
widely applicable to telecommunications, speech restoration, and high-fidelity
audio generation. Since the missing spectrum cannot be uniquely determined
from the observed signal, BWE must preserve available information while
generating plausible high-frequency content. In practice, input speech may
have different cutoff frequencies, whereas many existing systems are designed
for fixed source--target sampling-rate pairs.

Neural BWE has been studied in both the waveform and spectral domains.
TFiLM~\cite{tfilm} introduces recurrent feature-wise modulation for long-range
dependencies, while AFiLM~\cite{afilm} replaces recurrent modulation with
self-attention. NVSR~\cite{nvsr} predicts an extended mel spectrogram and
synthesizes the waveform using a pretrained neural vocoder. Spectral-domain
approaches model phase-related information: AERO~\cite{aero} predicts complex
STFT spectra, mdctGAN~\cite{mdctgan} performs adversarial learning in the MDCT
domain, and AP-BWE~\cite{ap_bwe} jointly extends amplitude and phase through
parallel streams. Fre-Painter~\cite{fre_painter} employs masked autoencoding
to improve robustness across input resolutions. Despite their effectiveness,
reconstruction-oriented models can produce over-smoothed high-frequency
content, while adversarial models must balance spectral fidelity and perceptual
realism within a single training stage.

Probabilistic generative models provide a natural way to model missing-frequency ambiguity. NU-Wave2~\cite{nuwave2} extends diffusion-based audio upsampling to multiple sampling rates, while UDM+~\cite{udm_plus} incorporates observations into reverse sampling for speech super-resolution. However, diffusion models typically require many evaluations. FlowHigh~\cite{flowhigh} introduces conditional flow matching~\cite{flow_matching} for audio super-resolution and achieves high-quality reconstruction with few-step sampling. A remaining challenge is to achieve accurate transport, one-step inference, and high perceptual quality without compromising the trajectory or high-frequency detail.

We propose \textbf{BDR-BWE}, a three-stage framework that decouples these
objectives. First, a cutoff-aware conditional flow bridge is learned in the
compressed complex STFT domain. Frequency-dependent stochasticity preserves
reliable observed components while focusing generation on the missing
spectrum. The learned multi-step trajectory is then compressed into few-step generators through progressive distillation~\cite{progressive_distillation}.
Finally, adversarial refinement improves high-frequency detail and harmonic
coherence. Experiments on VCTK demonstrate state-of-the-art log-spectral
distance across all evaluated input bandwidths, while evaluation on EARS
shows strong cross-dataset generalization. With only 1.80M parameters and one
or two network evaluations, BDR-BWE provides accurate and efficient
multi-bandwidth speech bandwidth extension.

\section{Proposed Method}
\label{sec:method}

\begin{figure*}[t]
\centering
\includegraphics[width=\linewidth]{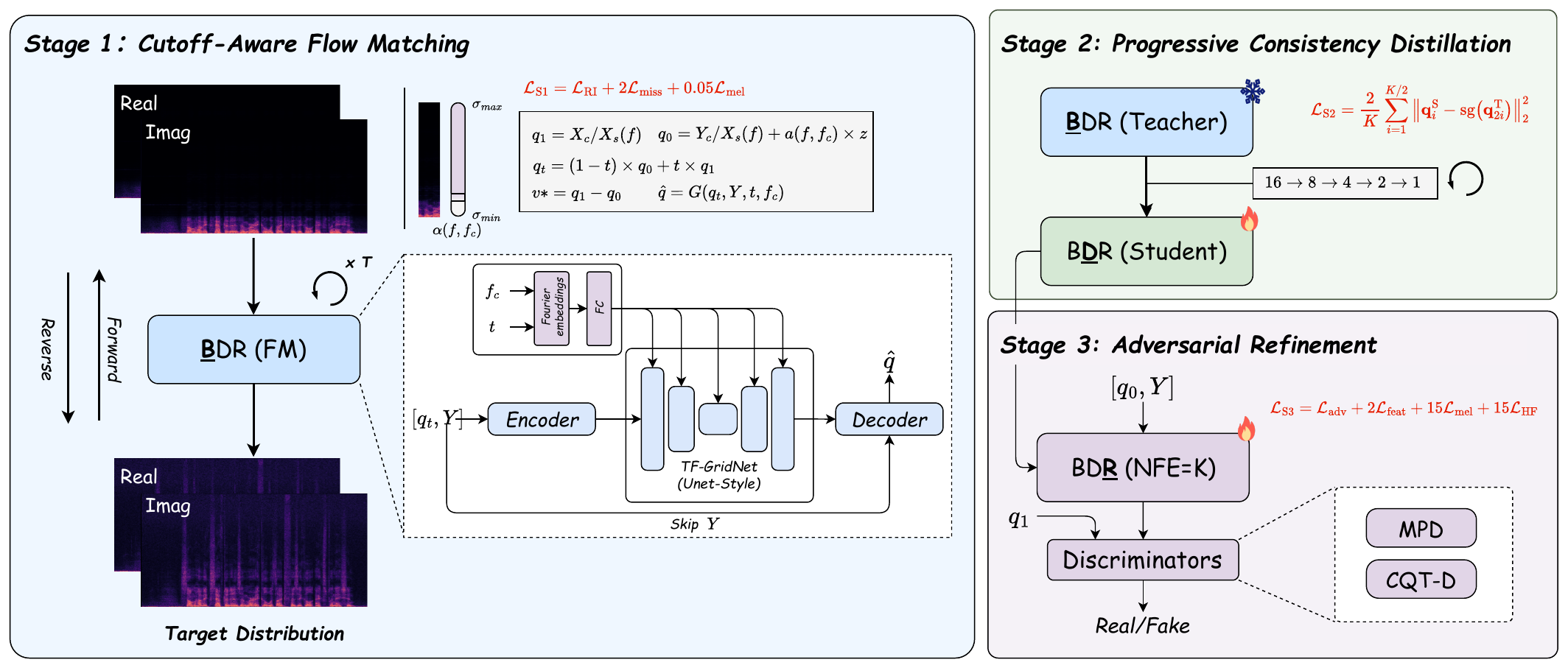}
\caption{Overview of BDR-BWE. Stage 1 learns a cutoff-aware flow bridge for
multi-bandwidth spectral reconstruction. Stage 2 progressively distills the
multi-step trajectory into few-step generators. Stage 3 further
refines the distilled generators with adversarial training to improve
high-frequency detail and perceptual quality. All stages share the same
complex-spectral generator.}
\label{fig:overview}
\end{figure*}

Given clean 48-kHz speech $\mathbf{x}\in\mathbb{R}^{N}$, we obtain a band-limited observation $\mathbf{y}=\mathcal{D}_{f_c}(\mathbf{x})$ with cutoff frequency $f_c$. We aim to restore full-band speech across varying input bandwidths using a single cutoff-conditioned model. As shown in Fig.~\ref{fig:overview}, BDR-BWE comprises three stages: cutoff-aware trajectory learning, progressive step reduction, and adversarial refinement.

\subsection{Stage 1: Cutoff-Aware Flow Bridge}
\label{sec:stage1}

Let $X=\operatorname{STFT}(\mathbf{x})$ and
$Y=\operatorname{STFT}(\mathbf{y})$. We apply complex power compression
$\mathcal{C}(Z)=0.3|Z|^{0.3}e^{j\angle Z}$ to preserve both magnitude and
phase while reducing the spectral dynamic range. The compressed spectra are
further normalized by a smoothed per-frequency RMS statistic $s_x(f)$
estimated from clean training speech, yielding
$\mathbf{q}_x=\mathcal{C}(X)/s_x(f)$ and
$\mathbf{q}_y=\mathcal{C}(Y)/s_x(f)$.

A conventional isotropic Gaussian source perturbs all frequency bins equally,
although frequencies below the cutoff already contain reliable observations.
We instead introduce a cutoff-aware noise profile $a(f,f_c)$ that smoothly
increases from a small value in the observed band to one in the missing band.
Let $w=\min(0.15f_c,3\,\mathrm{kHz})$,
$f_s=f_c-0.1w$, $f_e=f_c+0.9w$,
$u=(f-f_s)/(f_e-f_s)$, and $S(u)=3u^2-2u^3$. Then
\begin{equation}
a(f,f_c)=
\left\{
\begin{array}{ll}
0.05, & f<f_s,\\[2pt]
0.05+0.95S(u), & f_s\leq f\leq f_e,\\[2pt]
1, & f>f_e.
\end{array}
\right.
\label{eq:noise_profile}
\end{equation}

Given circular complex Gaussian noise
$\mathbf{z}\sim\mathcal{CN}(\mathbf{0},\mathbf{I})$, the bridge is defined as
\begin{equation}
\begin{aligned}
    \mathbf{q}_0
        &= \mathbf{q}_y+a(f,f_c)\odot\mathbf{z}, 
    \quad\mathbf{q}_1
        = \mathbf{q}_x,\\
    \mathbf{q}_t
        &= (1-t)\mathbf{q}_0+t\mathbf{q}_1,
        \quad t\sim\mathcal{U}(0,1).
\end{aligned}
\label{eq:bridge}
\end{equation}
This construction weakly perturbs the observed frequencies while retaining
full generative uncertainty in the missing band, thereby focusing the
transport on spectral completion.

The generator takes the real and imaginary components of
$(\mathbf{q}_t,\mathbf{q}_y)$ together with embeddings of $t$ and $f_c$, and
predicts the clean endpoint
$\widehat{\mathbf{q}}_1=
G_\theta(\mathbf{q}_t,\mathbf{q}_y,t,f_c)$.
Endpoint prediction has also been shown effective for flow-based audio
generation~\cite{flow2gan}. We train the model using a global complex-spectrum
reconstruction loss $\mathcal{L}_{\mathrm{RI}}$, a cutoff-aware reconstruction
loss $\mathcal{L}_{\mathrm{miss}}$ emphasizing the missing band, and a
multi-scale log-mel loss $\mathcal{L}_{\mathrm{mel}}$:
\begin{equation}
    \mathcal{L}_{\mathrm{S1}}
    =
    \mathcal{L}_{\mathrm{RI}}
    +2\mathcal{L}_{\mathrm{miss}}
    +0.05\mathcal{L}_{\mathrm{mel}}.
\label{eq:stage1_loss}
\end{equation}

Although trained with endpoint prediction, the model induces the local
velocity
$\widehat{\mathbf{v}}_\theta=
(\widehat{\mathbf{q}}_1-\mathbf{q}_t)/
\max(1-t,\epsilon)$, where $\epsilon=0.02$.
The resulting field is integrated with Euler updates during sampling.
The final spectrum is de-normalized, power-uncompressed, and converted to
waveform by inverse STFT. This formulation directly supervises the clean
complex spectrum while retaining a multi-step generative trajectory for
subsequent distillation.

\subsection{Stage 2: Progressive Consistency Distillation}
\label{sec:stage2}

While the Stage-1 model achieves accurate reconstruction with multiple solver
steps, iterative sampling limits inference efficiency. Progressive
distillation provides an effective way to repeatedly halve the sampling
steps of a trained generative model~\cite{progressive_distillation}, while
recent approaches such as Shortcut Models further demonstrate the potential
of one- and few-step generative transport~\cite{shortcut}. Motivated by these
advances, we progressively compress the learned BWE trajectory into one- and
two-step generators.

Starting from a $K$-step teacher, the student replaces every two teacher
updates with a single update. Teacher and student share the same stochastic
initialization $\mathbf{q}_0$ and use aligned uniform time grids. Let
$\mathbf{q}^{\mathrm{S}}_i$ denote the state after the $i$-th student update
and $\mathbf{q}^{\mathrm{T}}_{2i}$ the corresponding teacher state after two
updates. The distillation objective is
\begin{equation}
    \mathcal{L}_{\mathrm{S2}}
    =
    \frac{2}{K}
    \sum_{i=1}^{K/2}
    \left\|
        \mathbf{q}^{\mathrm{S}}_i
        -
        \operatorname{sg}
        \bigl(\mathbf{q}^{\mathrm{T}}_{2i}\bigr)
    \right\|_2^2,
\label{eq:distill_loss}
\end{equation}
where $\operatorname{sg}(\cdot)$ stops gradients through the teacher.

We progressively reduce the number of function evaluations (NFEs) following
$16\!\rightarrow\!8\!\rightarrow\!4\!\rightarrow\!2\!\rightarrow\!1$,
with each student initialized from its preceding teacher. No adversarial or
perceptual objective is introduced during distillation, allowing the student
to focus solely on reproducing the teacher trajectory. The resulting one- and
two-step models are retained for subsequent refinement.

\subsection{Stage 3: Adversarial Refinement}
\label{sec:stage3}

Compressing a multi-step trajectory into very few steps can smooth fine
spectral structures, particularly in the reconstructed high-frequency band.
Motivated by the effectiveness of adversarial training in neural
vocoders~\cite{bigvgan,vocos} and recent flow-based audio
generation~\cite{flow2gan}, we further refine the distilled generator using
adversarial objectives.

We employ a multi-period discriminator (MPD) for waveform periodicity and a
multi-scale sub-band constant-Q transform discriminator (CQT-D)
~\cite{cqt_discriminator} for complementary time--frequency supervision.
The latter provides frequency-dependent resolution that is well suited to
modeling harmonic structures. Both discriminators are trained with the hinge
adversarial objective. The generator combines adversarial loss
$\mathcal{L}_{\mathrm{adv}}$, feature matching
$\mathcal{L}_{\mathrm{feat}}$, multi-scale log-mel reconstruction
$\mathcal{L}_{\mathrm{mel}}$, and a cutoff-masked high-frequency
log-magnitude loss $\mathcal{L}_{\mathrm{HF}}$:
\begin{equation}
    \mathcal{L}_{\mathrm{S3}}
    =
    \mathcal{L}_{\mathrm{adv}}
    +2\mathcal{L}_{\mathrm{feat}}
    +15\mathcal{L}_{\mathrm{mel}}
    +15\mathcal{L}_{\mathrm{HF}}.
\label{eq:stage3_loss}
\end{equation}

The adversarial objective is gradually introduced at the beginning of
fine-tuning for stable optimization. Unlike a separate post-processing
module, adversarial refinement directly optimizes the fixed-NFE generator,
with gradients propagated through the complete few-step solver. This
allows the deployed generator to recover sharper high-frequency details and
more coherent harmonic structures without increasing the number of inference
steps.

\subsection{Shared Complex-Spectral Generator}
\label{sec:generator}

All three stages share the same compact complex-spectral generator, consisting of a spectral encoder, a U-Net-style TF-GridNet backbone, and a residual decoder. Given the real and imaginary components of $(\mathbf{q}_t,\mathbf{q}_y)$, the encoder increases the channel dimension while compressing the frequency axis to reduce the cost of subsequent time--frequency modeling. The backbone comprises five TF-GridNet blocks~\cite{tfgridnet} organized in a symmetric U-Net, where each block performs bidirectional frequency and temporal modeling followed by full-band temporal attention, with the original BiLSTMs replaced by lightweight BiGRUs. The flow time $t$ and cutoff frequency $f_c$ are encoded with Fourier embeddings and injected into the blocks to condition different flow states and input bandwidths. Finally, the decoder restores the frequency resolution and predicts a complex residual with respect to the observed spectrum $\mathbf{q}_y$, which is added to $\mathbf{q}_y$ to obtain the clean endpoint $\widehat{\mathbf{q}}_1$, effectively preserving observed spectral content while reconstructing the missing band.

\begin{table}[t]
\caption{Comparison with state-of-the-art BWE systems on VCTK. All systems
generate 48-kHz speech. Baselines are evaluated using their official
checkpoints under the same evaluation protocol.}
\label{tab:vctk_main_results}
\centering
\begingroup
\setlength{\tabcolsep}{2.1pt}
\resizebox{\linewidth}{!}{%
\begin{tabular}{@{}clccccc@{}}
\toprule
\textbf{Input SR} & \textbf{Method} & \textbf{LSD}$\downarrow$ &
\textbf{LSD-LF}$\downarrow$ & \textbf{LSD-HF}$\downarrow$ &
\textbf{ViSQOL}$\uparrow$ & \textbf{FAD}$\downarrow$ \\
\midrule
\multirow{6}{*}{8 kHz}
 & NU-Wave2~\cite{nuwave2}       & 1.487 & 0.521 & 1.608 & 1.712 & 1.478 \\
 & Fre-Painter~\cite{fre_painter} & 0.880 & 0.366 & 0.948 & 3.216 & 0.351 \\
 & AP-BWE~\cite{ap_bwe}          & 0.867 & 0.212 & 0.944 & 3.244 & 0.370 \\
 & FlowHigh~\cite{flowhigh}      & 0.820 & 0.203 & 0.892 & \textbf{3.389} & 0.207 \\
 & Ours (NFE=1)                  & \underline{0.755} & \underline{0.122} & \underline{0.825} & 3.284 & \textbf{0.135} \\
 & Ours (NFE=2)                  & \textbf{0.748} & \textbf{0.119} & \textbf{0.817} & \underline{3.309} & \underline{0.152} \\
\midrule
\multirow{6}{*}{12 kHz}
 & NU-Wave2~\cite{nuwave2}       & 1.313 & 0.555 & 1.477 & 1.914 & 1.073 \\
 & Fre-Painter~\cite{fre_painter} & 0.841 & 0.427 & 0.937 & 3.236 & 0.234 \\
 & AP-BWE~\cite{ap_bwe}          & 0.805 & 0.205 & 0.921 & 3.413 & 0.164 \\
 & FlowHigh~\cite{flowhigh}      & 0.763 & 0.222 & 0.869 & \textbf{3.505} & 0.055 \\
 & Ours (NFE=1)                  & \underline{0.690} & \underline{0.122} & \underline{0.792} & 3.427 & \underline{0.045} \\
 & Ours (NFE=2)                  & \textbf{0.685} & \textbf{0.120} & \textbf{0.787} & \underline{3.444} & \textbf{0.043} \\
\midrule
\multirow{6}{*}{16 kHz}
 & NU-Wave2~\cite{nuwave2}       & 1.206 & 0.579 & 1.412 & 2.145 & 0.961 \\
 & Fre-Painter~\cite{fre_painter} & 0.769 & 0.395 & 0.897 & 3.448 & 0.051 \\
 & AP-BWE~\cite{ap_bwe}          & 0.758 & 0.212 & 0.915 & 3.418 & 0.162 \\
 & FlowHigh~\cite{flowhigh}      & 0.715 & 0.237 & 0.857 & \textbf{3.660} & 0.044 \\
 & Ours (NFE=1)                  & \underline{0.632} & \underline{0.122} & \underline{0.768} & 3.636 & \textbf{0.027} \\
 & Ours (NFE=2)                  & \textbf{0.627} & \textbf{0.120} & \textbf{0.762} & \underline{3.651} & \underline{0.028} \\
\midrule
\multirow{6}{*}{24 kHz}
 & NU-Wave2~\cite{nuwave2}       & 1.036 & 0.611 & 1.318 & 2.814 & 1.198 \\
 & Fre-Painter~\cite{fre_painter} & 0.669 & 0.383 & 0.863 & 3.902 & 0.016 \\
 & AP-BWE~\cite{ap_bwe}          & 0.649 & 0.187 & 0.896 & 3.931 & 0.070 \\
 & FlowHigh~\cite{flowhigh}      & 0.612 & 0.239 & 0.829 & 4.212 & 0.010 \\
 & Ours (NFE=1)                  & \underline{0.524} & \underline{0.110} & \underline{0.732} & \underline{4.216} & \textbf{0.008} \\
 & Ours (NFE=2)                  & \textbf{0.519} & \textbf{0.108} & \textbf{0.725} & \textbf{4.217} & \underline{0.009} \\
\bottomrule
\end{tabular}%
}
\endgroup
\end{table}

\section{Experimental Setup}
\label{sec:experiments}

\subsection{Datasets and Bandwidth Degradation}
\label{sec:data}

\textbf{VCTK.}
We use VCTK~\cite{vctk}, containing approximately 44 hours of 48-kHz speech
from 108 English speakers. Following~\cite{nuwave2,fre_painter,flowhigh}, 100 speakers
are used for training and eight for evaluation. Band-limited inputs are
generated using a Chebyshev Type-I low-pass filter followed by downsampling
and resampling to 48 kHz. During training, the filter order is sampled from
2 to 10, and the input sampling rate is selected
from $\{4,6,\ldots,32\}$ kHz. For evaluation, we use an 8th-order filter and
report input rates of 8, 12, 16, and 24 kHz.

\textbf{EARS.}
We further evaluate out-of-domain generalization on eight speakers from
EARS~\cite{ears}, without using any EARS data for training or model selection.
The same evaluation degradation as VCTK is applied, and results are reported
for 4- and 8-kHz input conditions.

\subsection{Implementation Details}
\label{sec:implementation}

Training uses random one-second 48-kHz segments. We use a 1024-point STFT with
a 1024-sample Hann window and a 512-sample hop. The five U-Net GridBlocks have
channel dimensions $24/48/96/48/24$ and use bidirectional GRUs with 80 hidden
units. Gaussian Fourier embeddings of time and cutoff are injected into every
GridBlock. The generator contains 1.80M parameters and requires 16.8 GMACs per
second of audio for each network evaluation. Stage 1 is trained for 300k steps using AdamW with a peak learning rate of
$5\times10^{-4}$ and 5k warm-up steps. Each progressive distillation stage is
trained for 5k steps with a peak learning rate of $10^{-4}$. Stage 3 is
fine-tuned for 20k steps with generator and discriminator learning rates of
$2\times10^{-5}$ and $10^{-4}$, respectively. Adversarial losses are ramped
over the first 2k updates. We use EMA with a decay of 0.999 and conduct all
experiments in FP32. We use multi-scale log-mel resolutions, an MPD with periods
$\{2,3,5,7,11\}$, and a three-resolution CQT discriminator. The adversarial
and feature-matching objectives follow BigVGAN~\cite{bigvgan}.

\subsection{Evaluation Metrics}
\label{sec:metrics}

% We report log-spectral distance (LSD) using a 2048-point FFT with a 512-sample hop \cite{nuwave2, flowhigh}. LSD-LF and LSD-HF are computed below and above the input Nyquist frequency, respectively, measuring preservation of the observed spectrum and reconstruction of the missing high-frequency spectrum. Perceptual quality is evaluated using ViSQOL~\cite{visqol}. We additionally report Fréchet Audio Distance (FAD)~\cite{fad}, computed using features extracted by the 48-kHz EnCodec encoder~\cite{encodec}. FAD is measured between the clean and generated evaluation sets, with lower values indicating closer agreement between their feature distributions.

We report log-spectral distance (LSD) using a 2048-point FFT with a 512-sample hop~\cite{nuwave2,flowhigh}. LSD-LF and LSD-HF are computed below and above the input Nyquist frequency, respectively, to measure observed-band preservation and missing-band reconstruction. Perceptual quality is evaluated using ViSQOL~\cite{visqol}. We additionally report Fr'echet Audio Distance (FAD)~\cite{fad}, computed between the clean and generated evaluation sets using the \texttt{frechet-audio-distance}\footnote{\url{https://github.com/gudgud96/frechet-audio-distance}} implementation with embeddings extracted by the 48-kHz EnCodec encoder~\cite{encodec}. Lower FAD indicates closer distributional agreement between generated and real speech.

\section{Experimental Results}
\label{sec:results}

We compare BDR-BWE with representative flow- and GAN-based BWE systems, including NU-Wave2~\cite{nuwave2}, Fre-Painter~\cite{fre_painter}, AP-BWE~\cite{ap_bwe}, and FlowHigh~\cite{flowhigh}. All baselines are evaluated from their official checkpoints using the same degradation and evaluation protocol.

\begin{table}[t]
\caption{Comparison in the challenging 4-kHz-to-48-kHz setting.}
\label{tab:hardcase_results}
\centering
\begingroup
\setlength{\tabcolsep}{2.5pt}
\begin{tabular}{@{}lccccc@{}}
\toprule
\textbf{Method} & \textbf{LSD}$\downarrow$ & \textbf{ViSQOL}$\uparrow$ &
\textbf{Params} & \textbf{NFE} & \textbf{GMACs} \\
\midrule
NU-Wave2~\cite{nuwave2}        & 1.697 & 1.619 & 1.71M  & 8 & $27.7\times8$ \\
Fre-Painter~\cite{fre_painter} & 0.922 & 3.217 & 22.95M & 1 & $82.7/4.8$ \\
FlowHigh~\cite{flowhigh}       & 0.901 & \underline{3.230} & 49.40M & 1 & 25.5 \\
Ours (NFE=1)                   & \underline{0.831} & 3.224 & 1.80M & 1 & 16.8 \\
Ours (NFE=2)                   & \textbf{0.823} & \textbf{3.258} & 1.80M & 2 & $16.8\times2$ \\
\bottomrule
\end{tabular}
\endgroup
\end{table}

\subsection{Comparison with State-of-the-Art Methods}
\label{sec}

Table~\ref{tab:vctk_main_results} compares BDR-BWE with existing methods on the VCTK dataset at input sampling rates of 8, 12, 16, and 24 kHz. BDR-BWE achieves the best LSD, LSD-LF, and LSD-HF across all input bandwidths, indicating that it can better preserve the observed spectrum while more accurately reconstructing the missing high-frequency components. Compared with the strongest baseline, FlowHigh, the one-step BDR-BWE reduces LSD by 7.9\%, 9.6\%, 11.6\%, and 14.4\% at the four input bandwidths, respectively, with the advantage becoming more pronounced as the bandwidth-extension difficulty decreases. Although NFE=2 generally yields the best spectral results, NFE=1 remains very close, with a maximum LSD gap of only 0.007. In addition, the one-step model achieves the best FAD in three of the four settings while remaining competitive in ViSQOL, and further attains the best ViSQOL at the 24-kHz input setting. Overall, BDR-BWE can effectively handle multiple input bandwidths with a single cutoff-conditioned model while maintaining high spectral reconstruction accuracy and perceptual quality with one-step inference.

\subsection{Challenging Low-Bandwidth Setting}
\label{sec:results_hardcase}

We further consider the challenging 4-kHz-to-48-kHz setting, where most of the
target spectrum must be generated from a narrow observed band.
Table~\ref{tab:hardcase_results} shows that BDR-BWE remains effective under
this extreme degradation. Compared with FlowHigh, the one-step model reduces
LSD from 0.901 to 0.831 while using only 1.80M parameters and 16.8 GMACs per
second of audio. Increasing the NFE to two further improves LSD to 0.823 and
achieves the best ViSQOL of 3.258. Notably, the one-step model already
outperforms substantially larger competing systems in spectral reconstruction,
highlighting the favorable quality--efficiency trade-off of BDR-BWE.

\subsection{Out-of-Domain Generalization}
\label{sec:results_ood}

We evaluate cross-corpus generalization on eight unseen EARS speakers without any fine-tuning. As shown in Table~\ref{tab:ears_ood_results}, BDR-BWE achieves the best spectral and perceptual performance for both 4- and 8-kHz inputs. Compared with FlowHigh, the one-step model reduces LSD from 1.368 to 1.060 and from 1.319 to 0.940, respectively. The consistent improvements in both LSD-LF and LSD-HF further indicate strong generalization to unseen speakers and recording conditions.

\subsection{Ablation Study}
\label{sec:results_ablation}

Table~\ref{tab:three_stage_ablation} validates the effectiveness of the three-stage training strategy under the 8-kHz-to-48-kHz setting. Stage 1 achieves better generation quality with higher NFEs, reaching an LSD of 0.926 at 16 NFEs. Stage 2 progressively compresses the multi-step trajectory into a one-step model through progressive consistency distillation while largely preserving performance, yielding an LSD of 0.935. Stage 3 further reduces LSD to 0.755 through adversarial refinement, with clear improvements in ViSQOL and FAD. These results demonstrate the complementary roles of trajectory learning, progressive distillation, and adversarial refinement in balancing generation quality and one-step inference.

\begin{table}[t]
\caption{Out-of-domain evaluation on EARS. All systems generate 48-kHz speech.}
\label{tab:ears_ood_results}
\centering
\begingroup
\setlength{\tabcolsep}{2.1pt}
\begin{tabular}{@{}clcccc@{}}
\toprule
\textbf{Input SR} & \textbf{Method} & \textbf{LSD}$\downarrow$ &
\textbf{LSD-LF}$\downarrow$ & \textbf{LSD-HF}$\downarrow$ &
\textbf{ViSQOL}$\uparrow$ \\
\midrule
\multirow{5}{*}{4 kHz}
 & NU-Wave2~\cite{nuwave2}        & 2.471 & 0.487 & 2.575 & 2.023 \\
 & Fre-Painter~\cite{fre_painter} & 1.456 & 0.361 & 1.516 & 3.072 \\
 & FlowHigh~\cite{flowhigh}       & 1.368 & 0.205 & 1.426 & 3.158 \\
 & Ours (NFE=1)                   & \underline{1.060} & \underline{0.125} & \underline{1.106} & \underline{3.190} \\
 & Ours (NFE=2)                   & \textbf{1.059} & \textbf{0.119} & \textbf{1.104} & \textbf{3.197} \\
\midrule
\multirow{5}{*}{8 kHz}
 & NU-Wave2~\cite{nuwave2}        & 2.347 & 0.532 & 2.557 & 1.939 \\
 & Fre-Painter~\cite{fre_painter} & 1.358 & 0.421 & 1.474 & 3.165 \\
 & FlowHigh~\cite{flowhigh}       & 1.319 & 0.224 & 1.441 & 3.261 \\
 & Ours (NFE=1)                   & \textbf{0.940} & \underline{0.123} & \textbf{1.028} & \textbf{3.278} \\
 & Ours (NFE=2)                   & \underline{0.942} & \textbf{0.120} & \underline{1.030} & \underline{3.272} \\
\bottomrule
\end{tabular}
\endgroup
\end{table}

% \subsection{Ablation Study}
% \label{sec:results_ablation}

% Table~\ref{tab:three_stage_ablation} isolates the contribution of each
% training stage in the 8-kHz-to-48-kHz setting. The Stage-1 flow-matching
% model benefits substantially from iterative integration: increasing the
% NFE from one to 16 reduces LSD from 2.470 to 0.926. Stage 2 transfers this
% multi-step behavior to low-NFE students. Its one-step model reaches an LSD
% of 0.935, leaving only a small gap to the 16-step Stage-1 teacher, and its
% performance remains stable across the evaluated NFEs. Stage 3 then improves
% the one-step output from 0.935 to 0.755 LSD, from 2.954 to 3.284 ViSQOL, and
% from 0.268 to 0.135 FAD. The two-step final model gives the best spectral
% reconstruction and ViSQOL, whereas the one-step model achieves the best FAD.
% Together, the results verify the complementary roles of flow matching,
% consistency distillation, and adversarial refinement.

\begin{table}[t]
\caption{Ablation of the three-stage training framework for 8-kHz-to-48-kHz
BWE.}
\label{tab:three_stage_ablation}
\centering
\begingroup
\setlength{\tabcolsep}{2.1pt}
% \resizebox{\linewidth}{!}{%
\begin{tabular}{@{}lcccccc@{}}
\toprule
\textbf{Stage} & \textbf{NFE} & \textbf{LSD}$\downarrow$ &
\textbf{LSD-LF}$\downarrow$ & \textbf{LSD-HF}$\downarrow$ &
\textbf{ViSQOL}$\uparrow$ & \textbf{FAD}$\downarrow$ \\
\midrule
\multirow{5}{*}{\shortstack{Stage 1\\(FM)}}
 & 1  & 2.470 & 0.127 & 2.705 & 2.480 & 0.970 \\
 & 2  & 2.569 & 0.124 & 2.814 & 2.073 & 1.146 \\
 & 4  & 1.732 & 0.126 & 1.896 & 2.144 & 1.088 \\
 & 8  & 1.100 & 0.130 & 1.203 & 2.602 & 0.687 \\
 & \textbf{16} & \textbf{0.926} & \textbf{0.134} & \textbf{1.013} & \textbf{2.957} & \textbf{0.220} \\
\midrule
\multirow{4}{*}{\shortstack{Stage 2\\(PCD)}}
 & 16$\rightarrow$8 & 0.931 & 0.133 & \textbf{1.018} & 2.959 & 0.267 \\
  & 8$\rightarrow$4 & 0.935 & 0.133 & 1.022 & 2.949 & 0.281 \\
 & 4$\rightarrow$2 & 0.936 & 0.134 & 1.023 & 2.949 & 0.275 \\
 & 2$\rightarrow$1 & 0.935 & 0.140 & 1.022 & 2.954 & 0.268 \\

\midrule
\multirow{2}{*}{\shortstack{Stage 3\\(GAN)}}
 & 1 & \underline{0.755} & \underline{0.122} & \underline{0.825} & \underline{3.284} & \textbf{0.135} \\
 & 2 & \textbf{0.748} & \textbf{0.119} & \textbf{0.817} & \textbf{3.309} & \underline{0.152} \\
\bottomrule
\end{tabular}%
% }
\endgroup
\end{table}

\section{Conclusion}
\label{sec:conclusion}

We presented BDR-BWE, a three-stage framework for efficient multi-bandwidth
speech bandwidth extension. BDR-BWE combines cutoff-aware flow bridging,
progressive distillation, and adversarial refinement to preserve observed
spectral content while reconstructing missing frequencies with only one or
two network evaluations. Experiments on VCTK demonstrate state-of-the-art
spectral reconstruction across multiple input bandwidths, while evaluation
on EARS shows strong out-of-domain generalization. With only 1.80M parameters,
the one-step model achieves a favorable balance between reconstruction
quality, perceptual quality, and inference efficiency.

\bibliographystyle{IEEEbib}
\bibliography{refs}

@inproceedings{tfilm,
  title     = {Temporal {FiLM}: Capturing Long-Range Sequence Dependencies with Feature-Wise Modulations},
  author    = {Birnbaum, Sawyer and Kuleshov, Volodymyr and Enam, Zayd and Koh, Pang Wei W. and Ermon, Stefano},
  booktitle = {Advances in Neural Information Processing Systems},
  volume    = {32},
  year      = {2019},
  url       = {https://proceedings.neurips.cc/paper/2019/hash/2afc4dfb14e55c6face649a1d0c1025b-Abstract.html}
}

@inproceedings{afilm,
  title     = {Self-Attention for Audio Super-Resolution},
  author    = {Rakotonirina, Nathana{"e}l Carraz},
  booktitle = {2021 IEEE 31st International Workshop on Machine Learning for Signal Processing (MLSP)},
  pages     = {1--6},
  year      = {2021},
  doi       = {10.1109/MLSP52302.2021.9596082}
}

@inproceedings{nuwave2,
  title     = {{NU-Wave 2}: A General Neural Audio Upsampling Model for Various Sampling Rates},
  author    = {Han, Seungu and Lee, Junhyeok},
  booktitle = {Interspeech 2022},
  pages     = {4401--4405},
  year      = {2022},
  doi       = {10.21437/Interspeech.2022-45},
  issn      = {2958-1796}
}

@inproceedings{nvsr,
  title     = {Neural Vocoder is All You Need for Speech Super-Resolution},
  author    = {Liu, Haohe and Choi, Woosung and Liu, Xubo and Kong, Qiuqiang and Tian, Qiao and Wang, DeLiang},
  booktitle = {Interspeech 2022},
  pages     = {4227--4231},
  year      = {2022},
  doi       = {10.21437/Interspeech.2022-11017},
  issn      = {2958-1796}
}

@inproceedings{aero,
  title     = {{AERO}: Audio Super Resolution in the Spectral Domain},
  author    = {Mandel, Moshe and Tal, Or and Adi, Yossi},
  booktitle = {2023 IEEE International Conference on Acoustics, Speech and Signal Processing (ICASSP)},
  pages     = {1--5},
  year      = {2023},
  doi       = {10.1109/ICASSP49357.2023.10095382}
}

@inproceedings{udm_plus,
  title     = {Conditioning and Sampling in Variational Diffusion Models for Speech Super-Resolution},
  author    = {Yu, Chin-Yun and Yeh, Sung-Lin and Fazekas, Gy{"o}rgy and Tang, Hao},
  booktitle = {2023 IEEE International Conference on Acoustics, Speech and Signal Processing (ICASSP)},
  pages     = {1--5},
  year      = {2023},
  doi       = {10.1109/ICASSP49357.2023.10095103}
}

@inproceedings{mdctgan,
  title     = {{mdctGAN}: Taming Transformer-Based {GAN} for Speech Super-Resolution with Modified {DCT} Spectra},
  author    = {Shuai, Chenhao and Shi, Chaohua and Gan, Lu and Liu, Hongqing},
  booktitle = {Interspeech 2023},
  pages     = {5112--5116},
  year      = {2023},
  doi       = {10.21437/Interspeech.2023-113},
  issn      = {2958-1796}
}

@article{fre_painter,
  title   = {Audio Super-Resolution with Robust Speech Representation Learning of Masked Autoencoder},
  author  = {Kim, Seung-Bin and Lee, Sang-Hoon and Choi, Ha-Yeong and Lee, Seong-Whan},
  journal = {IEEE/ACM Transactions on Audio, Speech, and Language Processing},
  volume  = {32},
  pages   = {1012--1022},
  year    = {2024},
  doi     = {10.1109/TASLP.2023.3349053}
}

@article{ap_bwe,
  title   = {Towards High-Quality and Efficient Speech Bandwidth Extension with Parallel Amplitude and Phase Prediction},
  author  = {Lu, Ye-Xin and Ai, Yang and Du, Hui-Peng and Ling, Zhen-Hua},
  journal = {IEEE/ACM Transactions on Audio, Speech, and Language Processing},
  volume  = {33},
  pages   = {236--250},
  year    = {2025},
  doi     = {10.1109/TASLP.2024.3519881}
}

@inproceedings{flowhigh,
  title     = {{FLowHigh}: Towards Efficient and High-Quality Audio Super-Resolution with Single-Step Flow Matching},
  author    = {Yun, Jun-Hak and Kim, Seung-Bin and Lee, Seong-Whan},
  booktitle = {2025 IEEE International Conference on Acoustics, Speech and Signal Processing (ICASSP)},
  pages     = {1--5},
  year      = {2025},
  doi       = {10.1109/ICASSP49660.2025.10888772}
}

@inproceedings{flow_matching,
  title     = {Flow Matching for Generative Modeling},
  author    = {Lipman, Yaron and Chen, Ricky T. Q. and Ben-Hamu, Heli and Nickel, Maximilian and Le, Matthew},
  booktitle = {International Conference on Learning Representations},
  year      = {2023},
  url       = {https://openreview.net/forum?id=PqvMRDCJT9t}
}

@inproceedings{progressive_distillation,
  title     = {Progressive Distillation for Fast Sampling of Diffusion Models},
  author    = {Salimans, Tim and Ho, Jonathan},
  booktitle = {International Conference on Learning Representations},
  year      = {2022},
  url       = {https://openreview.net/forum?id=TIdIXIpzhoI}
}

@misc{vctk,
  title        = {{CSTR VCTK Corpus}: English Multi-speaker Corpus for {CSTR} Voice Cloning Toolkit (Version 0.92)},
  author       = {Yamagishi, Junichi and Veaux, Christophe and MacDonald, Kirsten},
  year         = {2019},
  publisher    = {University of Edinburgh, The Centre for Speech Technology Research},
  doi          = {10.7488/ds/2645}
}

@inproceedings{ears,
  title     = {{EARS}: An Anechoic Fullband Speech Dataset Benchmarked for Speech Enhancement and Dereverberation},
  author    = {Richter, Julius and Wu, Yi-Chiao and Krenn, Steven and Welker, Simon and Lay, Bunlong and Watanabe, Shinji and Richard, Alexander and Gerkmann, Timo},
  booktitle = {Interspeech 2024},
  year      = {2024},
  pages={4873--4877},
}

@inproceedings{visqol,
  title     = {{ViSQOL v3}: An Open Source Production Ready Objective Speech and Audio Metric},
  author    = {Chinen, Michael and Lim, Felicia S. C. and Skoglund, Jan and Gureev, Nikita and O'Gorman, Feargus and Hines, Andrew},
  booktitle = {2020 Twelfth International Conference on Quality of Multimedia Experience (QoMEX)},
  pages     = {1--6},
  year      = {2020},
  doi       = {10.1109/QoMEX48832.2020.9123150}
}

@inproceedings{fad,
  title     = {Fr\'{e}chet Audio Distance: A Reference-Free Metric for Evaluating Music Enhancement Algorithms},
  author    = {Kilgour, Kevin and Zuluaga, Mauricio and Roblek, Dominik and Sharifi, Matthew},
  booktitle = {Interspeech 2019},
  pages     = {2350--2354},
  year      = {2019},
  doi       = {10.21437/Interspeech.2019-2219}
}

@article{encodec,
  title   = {High Fidelity Neural Audio Compression},
  author  = {D\'{e}fossez, Alexandre and Copet, Jade and Synnaeve, Gabriel and Adi, Yossi},
  journal = {Transactions on Machine Learning Research},
  year    = {2023},
  url     = {https://openreview.net/forum?id=ivCd8z8zR2}
}

@inproceedings{bigvgan,
  title     = {{BigVGAN}: A Universal Neural Vocoder with Large-Scale Training},
  author    = {Lee, Sang-gil and Ping, Wei and Ginsburg, Boris and Catanzaro, Bryan and Yoon, Sungroh},
  booktitle = {International Conference on Learning Representations},
  year      = {2023},
  url       = {https://openreview.net/forum?id=iTtGCMDEzS_}
}

@inproceedings{shortcut,
  title     = {One Step Diffusion via Shortcut Models},
  author    = {Frans, Kevin and Hafner, Danijar and Levine, Sergey and Abbeel, Pieter},
  booktitle = {International Conference on Learning Representations},
  year      = {2025}
}

@inproceedings{vocos,
  title     = {Vocos: Closing the Gap between Time-Domain and Fourier-Based Neural Vocoders for High-Quality Audio Synthesis},
  author    = {Siuzdak, Hubert},
  booktitle = {International Conference on Learning Representations},
  year      = {2024}
}

@inproceedings{flow2gan,
  title     = {{Flow2GAN}: Hybrid Flow Matching and {GAN} with Multi-Resolution Network for Few-step High-Fidelity Audio Generation},
  author    = {Yao, Zengwei and Kang, Wei and Zhu, Han and Guo, Liyong and Ye, Lingxuan and Kuang, Fangjun and Zhuang, Weiji and Li, Zhaoqing and Han, Zhifeng and Lin, Long and Povey, Daniel},
  booktitle = {International Conference on Learning Representations},
  year      = {2026}
}

@inproceedings{cqt_discriminator,
  title     = {Multi-Scale Sub-Band Constant-Q Transform Discriminator for High-Fidelity Vocoder},
  author    = {Gu, Yicheng and Zhang, Xueyao and Xue, Liumeng and Wu, Zhizheng},
  booktitle = {IEEE International Conference on Acoustics, Speech and Signal Processing (ICASSP)},
  year      = {2024},
  pages     = {10616-10620}
}

@article{tfgridnet,
  author  = {Wang, Zhong-Qiu and Cornell, Samuele and Choi, Shukjae
             and Lee, Younglo and Kim, Byeong-Yeol and Watanabe, Shinji},
  title   = {{TF-GridNet}: Integrating Full- and Sub-Band Modeling
             for Speech Separation},
  journal = {IEEE/ACM Transactions on Audio, Speech, and Language Processing},
  volume  = {31},
  pages   = {3221--3236},
  year    = {2023},
  doi     = {10.1109/TASLP.2023.3304482}
}

\end{document}